# Color-Blind Image Sensors: Towards Digital Twin of Human Retina

**Authors**
Yushan Meng[1]*, Bryce Widdicombe[1], Dechuan Sun[1], Paul Beckett[1], Peter van Wijngaarden[2,3], Efstratios Skafidas[1], Ampalavanapillai Nirmalathas[1], Ranjith Unnithan[1]*

**Affiliations**
[1]Department of Electrical and Electronic Engineering, Faculty of Engineering and Information Technology, The University of Melbourne. Parkville, VIC, 3010, Australia.
[2]Centre for Eye Research Australia, Royal Victorian Eye and Ear Hospital, East Melbourne.
[3]Ophthalmology, University of Melbourne Department of Surgery.

*Ranjith Unnithan: r.ranjith@unimelb.edu.au
*Yushan Meng: menym@student.unimelb.edu.au

**Abstract**
The human retina contains a complex arrangement of photoreceptors that convert light into visual information. Conventional image sensors mimic the trichromacy of the retina using periodic filter mosaics responsive to three primary colors. However, this is, at best, an approximation, as an actual retina exhibits a quasi-random spatial distribution of light-sensitive rod and cone photoreceptors, where the ratio of rods to cones and their concentrations vary across the retina. Hence, the periodic mosaics are limited to accurately simulate the properties of the eye. Here, we present an image sensor with similar distribution, spacing, ratios and spectral characteristics of an actual foveal mosaic for emulating eye-like sampling and mimicking color blindness. To perform image reconstruction, we use a fully convolutional U-Net neural network adopting the concept of receptive fields in the retinal circuitry. Our research will enable the development of digital twin of a retina to further understand color vision deficiencies.

**Keywords**
Image sensor, color blindness, human retina, bio-inspiration.

**Introduction**
Human color perception begins with the stimulation of light-sensitive cone and rod photoreceptor cells, which are quasi-randomly distributed in the retina (Parry, 2015). Cone cells, located in the central region of the retina known as the fovea, are crucial for color vision and the perception of fine visual details, functioning optimally under bright light (photopic) conditions. In contrast, rod cells exhibit a broader spectral response and operate under low to medium light (scotopic) conditions. These cells are more concentrated in the peripheral areas of the retina beyond the fovea and significantly outnumber the cone cells (Curcio et al., 1990; Wells-Gray et al., 2016). In humans with normal color vision, three classes of retinal cone cells, namely Short (S), Medium (M), and Long (L) wavelength sensitive cones, mediate nearly all color vision. These cones exhibit peak spectral responses at wavelengths of approximately 420 nm, 530 nm, and 558 nm, respectively (Bowmaker and Dartnall, 1980). The color spectra of S, M and L cones are broad and overlap considerably, with peak spectral responses influenced by individual physiology and genetics (Deeb, 2006; Webster, 2015).

Conventional color image sensors mimic trichromatic human vision by using a color filter array (CFA) comprising the primary colors red, green and blue (RGB) over monochromatic pixels, followed by an image processor to reconstruct full-color images (Amba et al., 2016; He et al., 2021, 2020; Lukac and Plataniotis, 2005). The CFA is usually arranged in a periodic pattern. The most widely used of these, the Bayer mosaic, arranges RGB color filters in a 2×2 periodic pattern consisting of a quarter red, a quarter blue, and half green filters (Amba et al., 2016; He et al., 2021, 2020). In contrast to this periodic CFA, human retinal cones exhibit a quasi-random spatial distribution, which varies among individuals (Ahnelt, 1998; Garrigan et al., 2010). Furthermore, even though the human eye is more sensitive to green light, the foveal mosaic contains a larger population of L cones on average (Brainard, 2015). It has also been shown that the distribution of S cones, which typically constitute less than 10% of the total cone population (Brainard, 2019; Roorda et al., 2001), is essentially random, whereas both the M and L cone distributions exhibit a slight tendency towards clumping (Hofer et al., 2005; Roorda et al., 2001; Roorda and Williams, 1999). Moreover, the spectral responses of cone cells differ significantly from those used in artificial color spaces (Brainard, 2015; Rowlands, 2020; Stockman and Brainard, 2015). The close spectral sensitivities of M and L cones enable fine discrimination in the red-green region of the human vision color space (Mollon, 1989). On the other hand, the absence of one or more of the cone types, or the diminished spectral sensitivities of their photo-pigments, can cause various forms of color vision deficiency, collectively known as color blindness.

Bio-mimetic design bridges the gap between biological systems and technical innovations to achieve new functions and properties that would be otherwise unattainable through conventional engineering approaches (Hashemi Farzaneh and Lindemann, 2019; M. S. Kim et al., 2023; Posch, 2012; Sarkar and Theuwissen, 2013; Yu et al., 2013). One example of bio-mimetic design, eye-inspired vision systems, have received increased attention recently. Example proposals to date include artificial vision systems mimicking feline and cuttlefish's eyes for better object detection under variable lighting conditions (Kim et al., 2024; M. Kim et al., 2023), curved photodetector arrays inspired by the retina in single-chambered eyes (Gu et al., 2020; Jung et al., 2011; Rao et al., 2021), and CMOS image sensor with periodic checker pattern rod- and cone-like color pixels (Kawada et al., 2009).

Here, we demonstrate for the first time an eye-inspired image sensor based on a color mosaic that mimics the complex distribution of human photoreceptors to create a new sensor technology for studying color vision deficiencies. As illustrated in Fig. 1, the proposed system is influenced by the multilayer topology of the human visual system. While the eye contains three classes of cone cells together with rod cells in the retina, our sensor comprises a monochrome image sensor for photodetection overlaid with a quasi-random color filter array (Q-CFA). The Q-CFA represents a realistic approximation of both the trichromacy and spectral responses of the cones (L: Long, M: Middle, and S: Short wavelength sensitive), as well as their distribution in the fovea. Rods are represented by a monochrome (M) channel in the Q-CFA. We realized the eye-inspired image sensor by precisely aligning and integrating the Q-CFA onto a monochrome image sensor and incorporating a fully convolutional neural network (CNN) that is roughly analogous to the multilayered retinal cells and subsequent neuronal signal processing in the visual cortex. The CNN combines the necessary demosaicing process with a raw RGBM to sRGB color space conversion which allows the sensor to operate as an 'eye-like' system, capable of capturing and reconstructing full-color images. A potential application of this technology lies in the creation of personalized mosaics derived from individual retinal cone distributions and cone sensitivity functions. This capability holds promise for advancing our understanding of color vision deficiencies by facilitating the development of a digital twin of human retina. To demonstrate the feasibility of this approach, we created a color-blind image sensor by substituting the Q-CFA with a protanopic color filter array (P-CFA) to generate images mimicking the view of people with protanopia/protanomaly. Various Ishihara plates were used to validate the output images, which have been proved to exhibit desired colors and patterns as recorded in the Ishihara book.

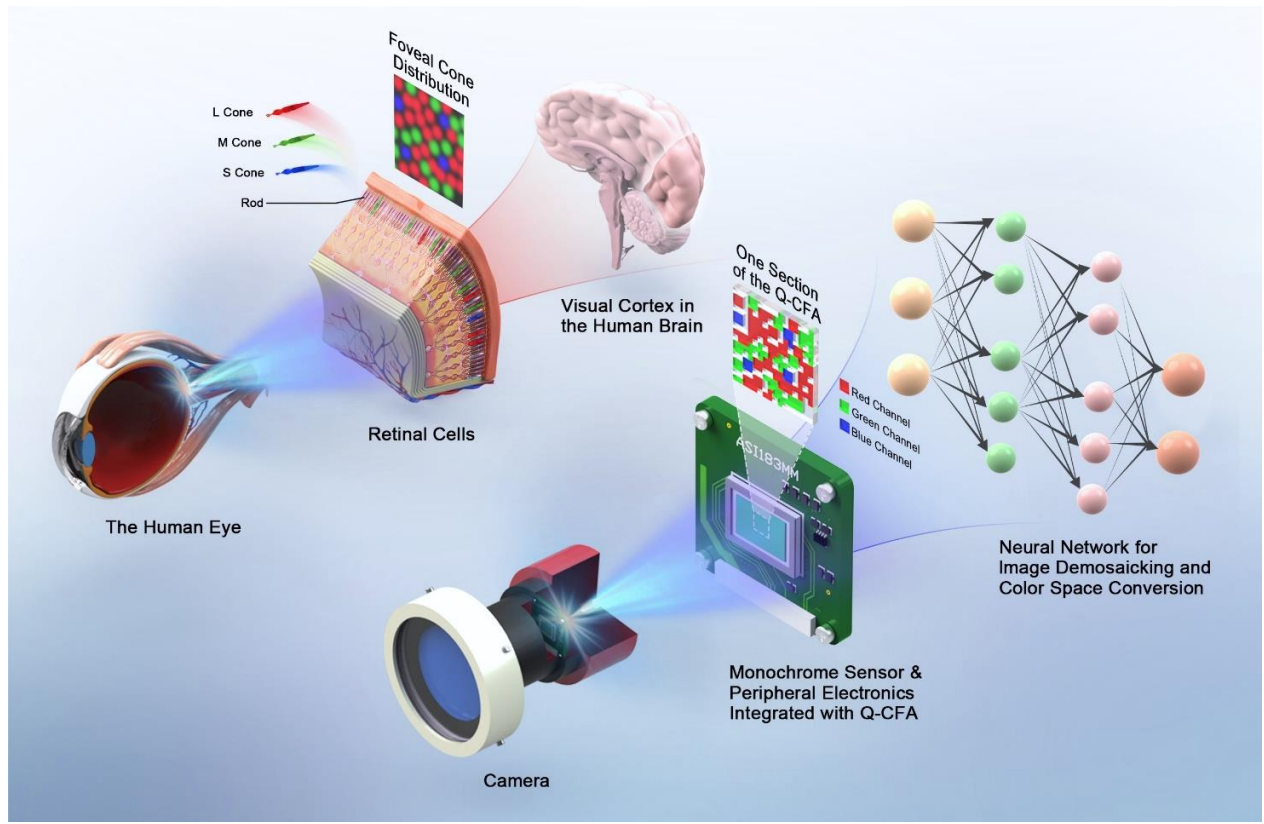

**Fig. 1. Color perception process of humans vs. the eye-inspired image sensor.** Top row from left to right: structure of the human eye, one section of the retinal cell hierarchy, example of foveal cone distribution (red, green, and blue indicate L, M, and S cones respectively), and visual cortex in the human brain; Bottom row from left to right: the digital camera with adaptable lens, the monochrome sensor integrated with the Q-CFA and peripheral electronics of the sensor, one section of the Q-CFA, and schematic representation of the neural network for image demosaicing and color space conversion.

## Results

### The quasi-random color filter array (Q-CFA) for eye-inspired image sensor

The first step towards making an image sensor that mimics the human eye, was to create a quasi-random color filter array that emulates the spatial distribution and spectral sensitivities of retinal cone cells. As shown in Fig. 1, the Q-CFA consists of three spectral bands corresponding to the three classes of cone photoreceptors in the human retina. To mimic the quasi-random distribution of the cone photoreceptors, we used a real retinal patch comprising 492 L, 236 M, and 39 S cones (size: 24×24 arcmin), captured at a 1.5° temporal eccentricity from the fovea (Sabesan et al., 2015). This patch was converted into a vector format, and the monochrome-to-colored pixel ratio after conversion has been validated comparing with the original rod-to-cone ratio in the retinal patch (please refer to supplementary information Section 1). It was then repeated to cover the 5496×3672 resolution of the IMX183 monochrome image sensor. Given the constraints of fabrication precision and pixel alignment, each filter pixel was designed to be 4.8μm in size, covering a 2×2-pixel block on the monochrome sensor (pixel size: 2.4μm). The complete color filter mosaic was separated into individual R/G/B channels and printed onto three mask plates for the Q-CFA fabrication. Then the Q-CFA was aligned and integrated onto the monochrome pixel array, followed by image demosaicing and color space conversion using a neural network. The simulation and fabrication details of the filter geometries are explained in Section 2 in the supplementary information. The Q-CFA samples images differently from periodic color filter arrays, exhibiting demosaicing advantages especially when the images have high frequency spatial components (please refer to supplementary information Section 3).

The optical responses of the three colored channels of the Q-CFA were measured using hyperspectral imaging on a Cytoviva Hyperspectral Microscope, and the results exhibited the desired performance (Fig. 2C). The peak transmission rates of the three channels exceed 30%, and their spectral sensitivity characteristics are similar to those of physical cones. Specifically, the green and red channels have wider bandwidths than the blue channel with considerable overlap between the green and red filter responses. We choose to shift the peak of the red filter slightly towards longer wavelength to facilitate the development of compatible image processing algorithms for the fabricated Q-CFA and to minimize the initial complexity of the neural network. The images of the fabricated mosaic in transmission mode using the Cytoviva Hyperspectral Microscope (Fig. 2B) and reflection mode using an Olympus BX53M (Fig. 2D) show that the pixels of the three colored channels on the filter mosaic are well-aligned and form a quasi-random distribution over the surface of the array.

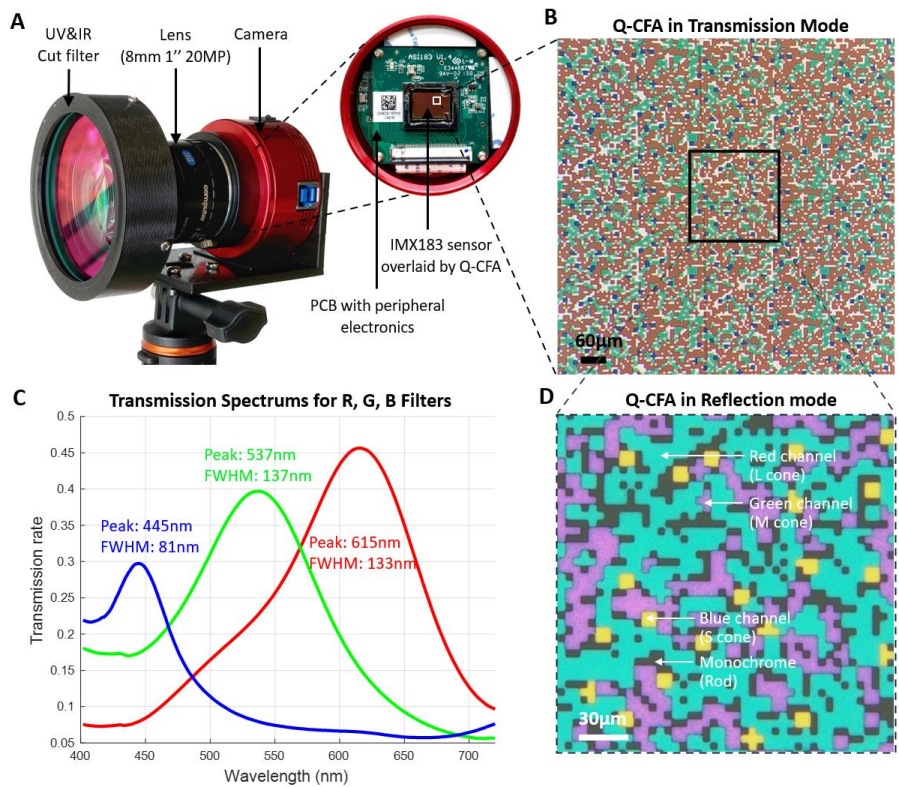

**Fig. 2. The eye camera and the quasi-random color filter array (Q-CFA).** (**A**) Hardware overview of the eye camera. (**B**) Optical picture of the color filter mosaic in transmission mode (scale bar: 60μm, the colors red, green, and blue correspond to the three respective channels). (**C**) Measured transmission spectra of the three channels (red, green, and blue) of the fabricated color filter mosaic. (**D**) Optical picture of the color filter mosaic in reflection mode (scale bar: 30μm, the colors cyan, magenta, and yellow correspond to channels red, green, and blue respectively).

**Full-color image recovery using a deep neural network**

We explored a fully convolutional neural network (CNN) and hybrid training set generation method to process the raw red, green, blue, monochrome (RGBM) mosaiced Q-CFA data into full-color images in a standard color space, sRGB. Although they are not direct equivalents, CNNs are inspired by the human visual system, adopting the concept of hierarchical visual data processing and pattern recognition through receptive fields of ganglion cells whilst incorporating modifications for computational efficiency. CFA demosaicing using convolutional neural networks has been previously demonstrated (Kurniawan et al., 2022; Shopovska et al., 2018). Our proposed one-time trained neural network aims to combine the demosaicing process with a raw RGBM to sRGB color space conversion, enabling simplified and "one-click" real-time processing of data captured directly from the hardware.

Prior to developing the convolutional neural network and color space converter, we first implemented a pixel mapping between the sensor and filters, which was used to create a mask to split captured images into RGBM bands. As shown in Fig. 3A, a ring of 12 RGB LEDs (WS2812B, Worldsemi co.) was placed above the sensor module within an enclosed box with a white acrylic diffuser between the LEDs and the lens. Images of the LEDs set to white (red, green, and blue all illuminated), red, green, and blue were captured. Then the red, green, and blue images were normalized against the white image, and a custom thresholding algorithm was used to generate an estimated filter pixel mask. This estimated mask served as a reference to manually align the photolithography fabrication mask on the pixel array, which could then be extrapolated across the full sensor. Fig. 3B demonstrates the mean values of each band (RGBM) of the eye-inspired image sensor when exposed to the red, green, and blue LEDs. The separation between the RGBM bands shows that the pixel mask correctly splits the image while also demonstrating the overlapping spectral responses of the filters. As a further validation, an estimate of the LED emission spectra was multiplied by the filter responses and sensor quantum efficiency and then integrated to estimate the energy captured by each band of the eye-inspired image sensor (Fig. 3C).

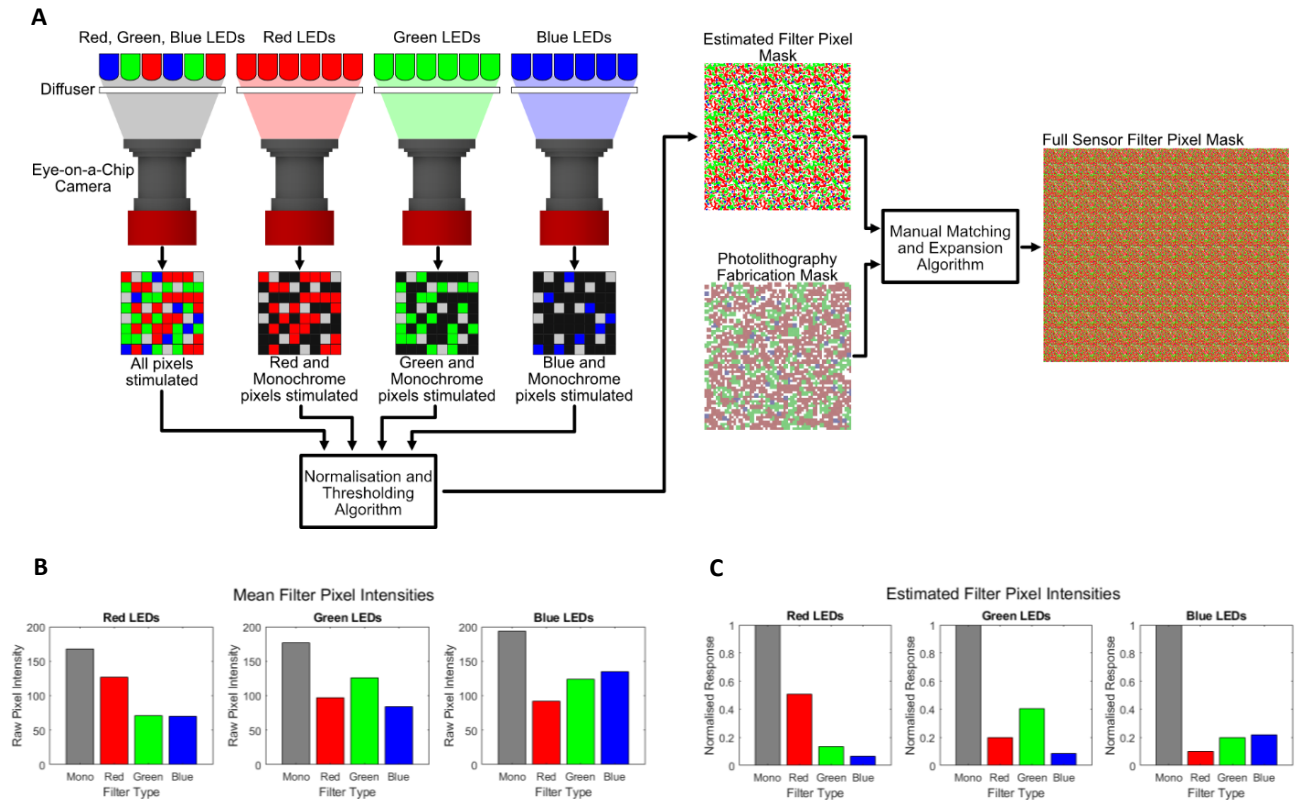

**Fig. 3. Pixel mask generation and validation of the pixel mask correctly splitting the image.** (**A**) Schematic illustration of the mapping between pixels on the sensor and filters, and the pixel mask generation process. (**B**) Mean filter pixel intensities of each band (RGBM) of the eye-inspired image sensor when exposed to red, green, and blue LEDs. (**C**) Estimated filter pixel intensities (normalized against the monochrome channel) captured by each band of the eye-inspired image sensor when exposed to red, green, and blue LEDs.

The traditional method for training a hardware image-to-image regression network is to mount the test camera next to a reference camera and capture images from both cameras simultaneously, which then need to be aligned to form a training dataset. As each filter in the Q-CFA on the eye-inspired image sensor covers a 2×2-pixel area, this alignment would need an accuracy of ±1 pixel to avoid misalignment artifacts. To prevent this issue, a custom-built hybrid synthetic/hardware training set was used comprising a color space converter and a mosaic generator. Samples of sRGB colors were imaged by the eye-inspired image sensor, and the corresponding raw averaged RGBM values were recorded. These pairs serve as inputs to train a regression model, which is used to convert sRGB data into simulated raw RGBM values. The visual representation of this color transformation is demonstrated in Fig. S4. To generate the synthetic neural network training set, sRGB images from various cameras were converted into RGBM images using the regression model. These converted images were then split into patches and each patch was sampled by either a random patch of the fabricated CFA or a randomly generated CFA pattern, with a 50/50 probability. The rationale behind the random CFA is to allow the network to generalize to other CFA patterns (evidence demonstrated in supplementary information Section 5), enabling study into the effects of color vision deficiencies.

A fully convolutional U-Net network architecture has been employed, with input size, encoder depths and filter sizes experimentally tuned (Fig. S6). A hybrid training dataset comprising 5491 samples was generated and randomly split into five sub-sets: four training and one validation. The network was trained in MATLAB using the ADAM (Adaptive Moment Estimation) optimization algorithm for 500 epochs, with RMSE (Root Mean Squared Error) used for the validation accuracy metric. The checkpoint with the lowest validation RMSE of 12.866, corresponding to epoch number 300 was exported into ONNX (Open Neural Network Exchange) format for implementation. A custom Python-based graphical user interface was developed to interface directly with the hardware, enabling automatic exposure setting, non-uniformity correction, real-time neural network inferencing and data collection.

We used the 24-patch Macbeth ColorChecker® board and Ishihara plates to verify the quality of the reconstructed colors. The Ishihara test was one of the earliest developed and is still the most widely used screening test for red-green color deficiency. Although early in its history some had argued that it was useful only as a rough screening test (Hardy et al., 1945), over time it has come to be widely accepted as a valid and accurate test for detecting red-green color vision deficiencies. In particular, it can be used to differentiate effectively between protan (red blindness) and deutan (green blindness) defects, which are the most common types of color vision deficiencies (Vision, 1981). Various studies have shown that the sensitivity of the Ishihara transformation and vanishing plates can reach up to 99% for protan/deutan defects (Birch, 1997; Birch and McKeever, 1993). Fig. 4A demonstrates the Ishihara plates used to test our trained neural network, while Fig. 4C shows that the neural network is able to reconstruct full-color images with satisfactory color accuracy (for a qualitative analysis, please refer to supplementary information Section 6). Notably, our algorithm is capable of preserving the hidden pattern of Ishihara plates. For example, in plate No. 21, the hidden pattern is encoded via the blue channel (Fig. 4 D4) to enable red-green deficiency detection. As shown in Fig. 4 E4, when employing our eye-inspired image sensor, the encoded "73" could be perceived after the original chart was mosaiced by the eye-inspired image sensor and demosaiced via the neural network.

**A color-blind image sensor based on a protanopic color filter array (P-CFA)**

Color vision deficiencies result from abnormalities in retinal cones, exhibiting as variations in cone distributions and spectral sensitivities (Brainard, 2015; Sabesan et al., 2015; Salih et al., 2020; Stockman and Brainard, 2015; Zhang et al., 2021). To demonstrate the concept of color-blind image sensors, we have built a new protanopic image sensor by implementing a color filter array that simulates a protanopic (caused by absent or defective L cones) foveal mosaic with spectral responses mimicking that of S and M cones. The protanopic foveal mosaic was designed by utilizing the ISETBio toolbox (Cottaris et al., 2019) with key design parameters outlined in Table 1. The cone sizes and spacings were further optimized based on real human photoreceptor data shown in previous studies (Curcio et al., 1990). The resulting complete filter array approximates a full fovea mosaic with a size of 7.4836×5 degrees, rather than a repeated sectional retinal patch. The two spectral bands of the protanopic color filter array (P-CFA) were designed and fabricated using a similar procedure as the Q-CFA, with the detailed design specifications outlined in supplementary information Section 2. Once fabricated, the P-CFA was integrated onto the same monochrome camera module (IMX183).

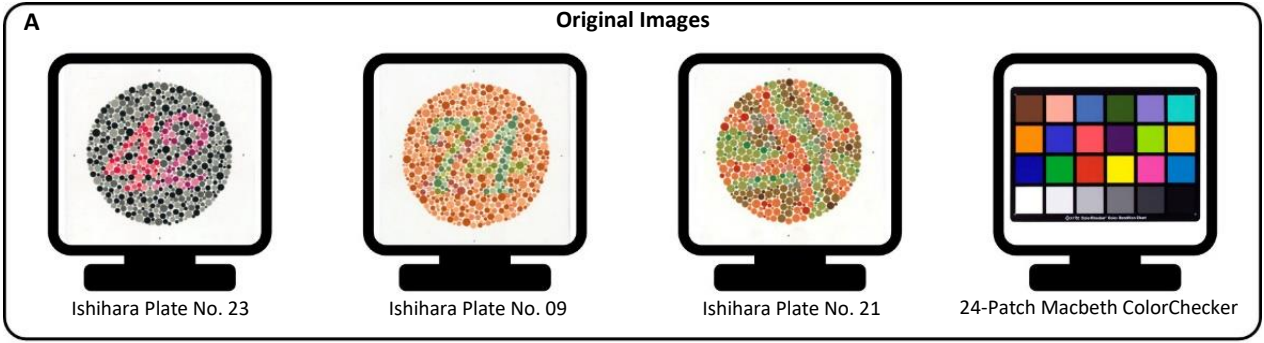

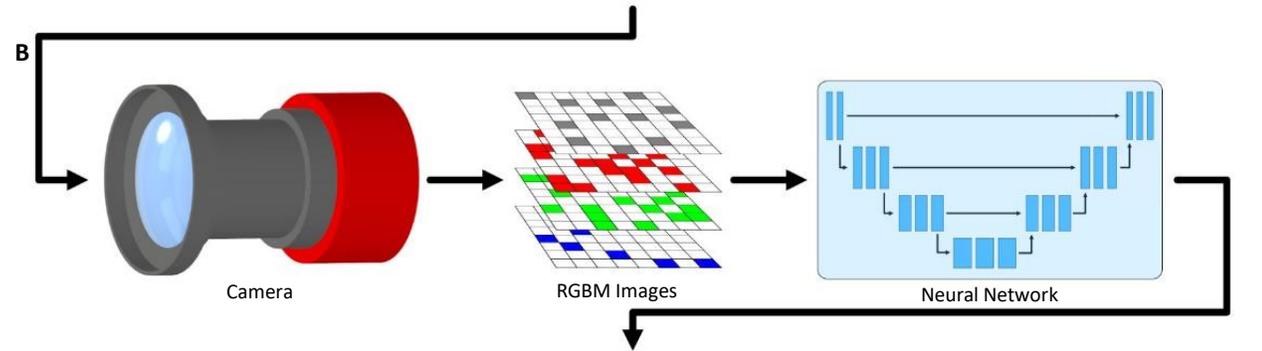

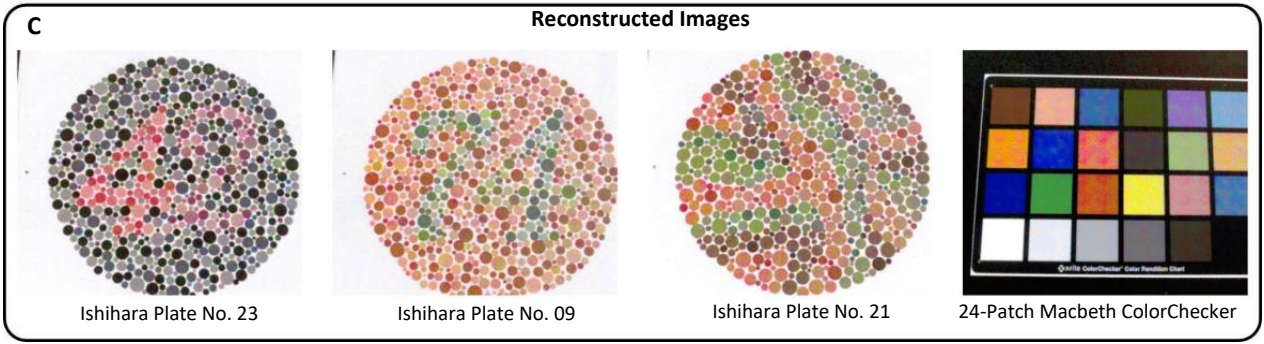

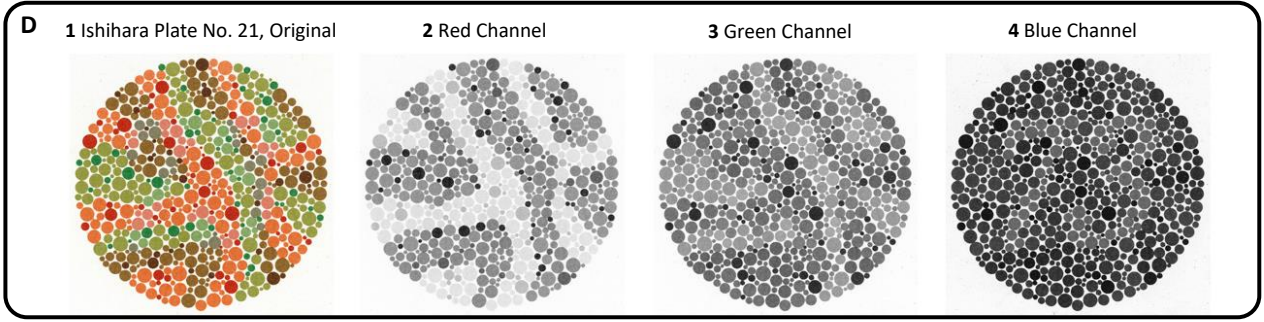

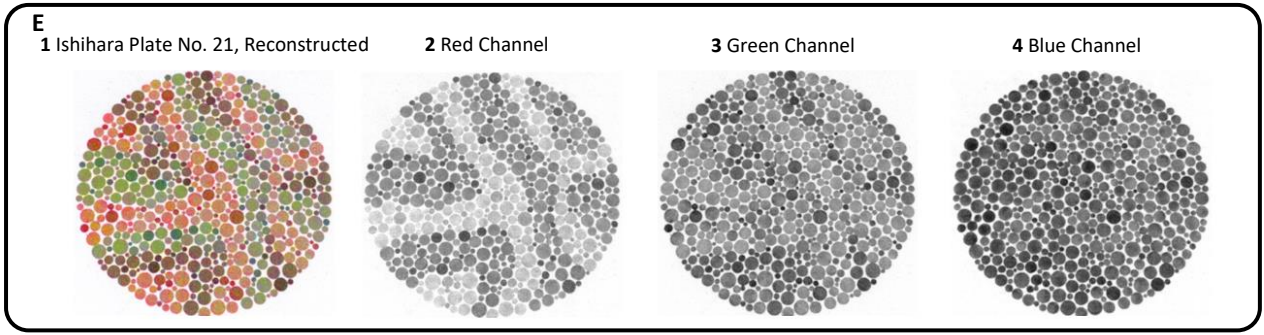

**Fig. 4. The eye-inspired image sensor and image reconstruction results.** (**A**) Original images of Ishihara plates No. 23, No. 09, No. 21, and the 24-patch Macbeth ColorChecker® (shown on computer screen). (**B**) Schematic illustration of the eye-inspired image sensor operating process. (**C**) Reconstructed images of Ishihara plates No. 23, No. 09, No. 21, and the 24-patch Macbeth ColorChecker® through the neural network. (**D**) Original image of Ishihara plate No. 21 and its red, green, and blue channel illustrations (the blue channel illustration shows the hidden pattern "73"). (**E**) Reconstructed image of Ishihara plate No. 21 and its red, green, and blue channel illustrations (the blue channel illustration shows the hidden pattern "73").

**Table 1.** Key design parameters for generating protanopic foveal mosaic

| Center of mosaic degrees | (0, 0) |
|---|---|
| Size degrees (width and height of the mosaic) | (7.4836, 5) |
| S-cone free region degrees | 0.15 |
| Cone population ratio L:M:S | 0:0.9:0.1 |

Images of the fabricated P-CFA in transmission and reflection modes are shown (Fig. 5A, 5B, 5D, 5E) to illustrate the growing cone size and spacing as eccentricity increases. Fig. 5C demonstrates the measured spectra of the two bands, which are further combined with the sensor quantum efficiency and UV-IR cut filter response, then normalized for comparison with the cone sensitivity functions. As shown in Fig. 5F, although the band 2 spectrum overlaps significantly with both the L and M cone sensitivities, its geometric center is closer to that of M cone along the wavelength axis, and it is clear that the band 2 spectrum has reduced intensity in the red region. Moreover, we have generated 100000 spectra with various shapes and peaks using a random spectrum generator. For each spectrum, we calculated the trapezoidal integrals of the spectrum multiplied with the L cone sensitivity, M cone sensitivity, and band 2 response respectively. These integrals were then normalized against their corresponding white response, which is the maximum energy that the specific cone or filter band can capture. By calculating the correlation between the L cone, M cone, and the filter band, we observed a slightly higher correlation (0.990978) between the M cone and the filter band than that between the L cone and the filter band (0.990753), with the correlation between the L and M cone being 0.971019. Therefore, we conclude that the new image sensor exhibits characteristics resembling a color-blindness that confuses red with green colors equally. Additionally, the sensor displays a reduction in red intensity and a geometric center that is more closely aligned with the M cone. These features collectively match the symptoms of protanopia (absent L cones) and protanomaly (defective L cones), making the sensor a protanopic camera.

Images captured by the newly integrated protanopic camera were utilized to train the neural network. The same network architecture and color transformation method (same method with different data, visual representation shown in Fig. S4) were used, with minor enhancements made to the training dataset generator. Specifically, the probability distribution for generating the training dataset was modified, using 60% random patches from the P-CFA and 40% randomly generated patterns. The random patterns were further refined by employing a blue noise distribution, which has been shown to best match the retinal cone distribution (Lanaro et al., 2020). Furthermore, the training dataset generator included random scaling and rotation of the input RGB images as data augmentation to prevent overfitting.

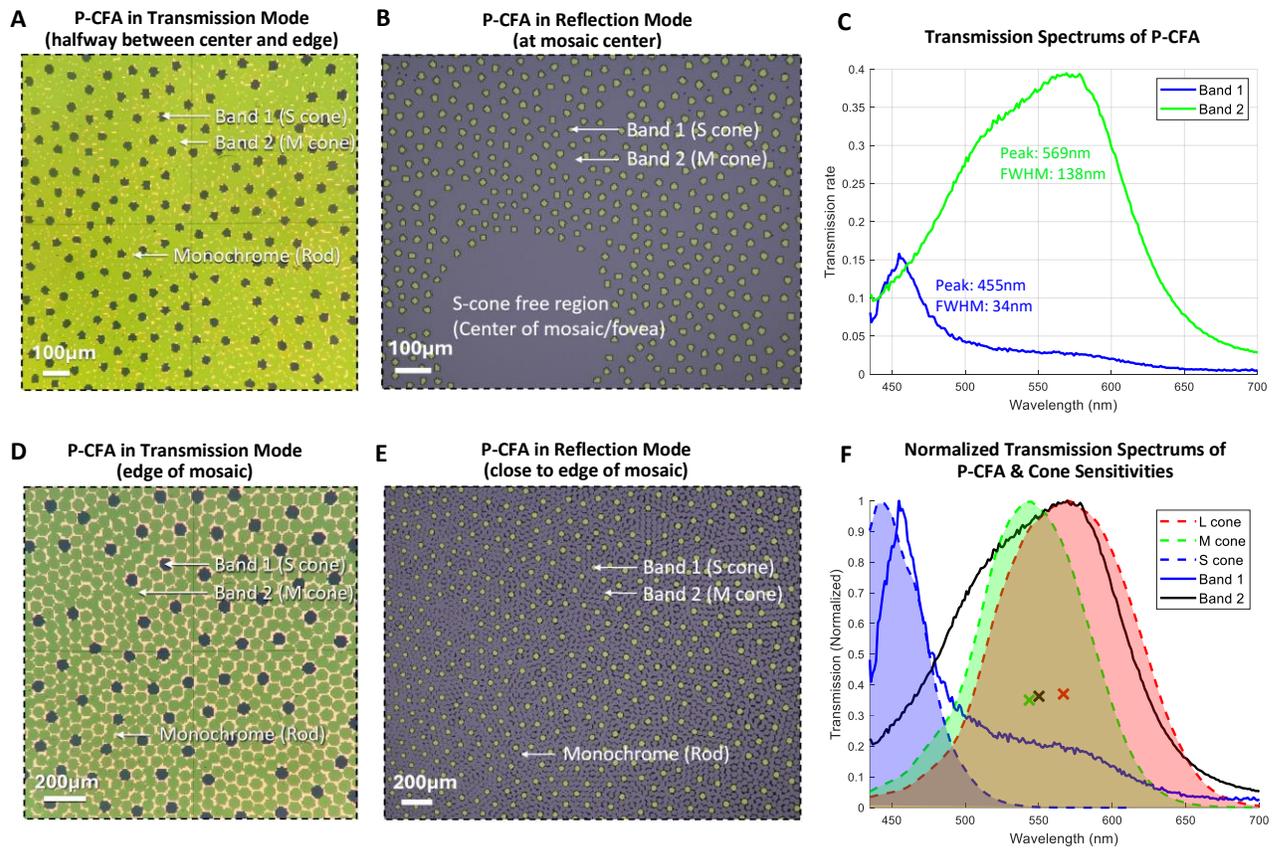

**Fig. 5. The protanopic color filter array (P-CFA).** (**A**) Optical picture of the color filter mosaic in transmission mode in halfway between the center and the edge of the P-CFA (scale bar: 100μm, the color blue and greenish yellow correspond to the two bands mimicking S and M cones). (**B**) Optical picture of the color filter mosaic in reflection mode at the center of the P-CFA (scale bar: 100μm, the color gold and purple correspond to the two bands mimicking S and M cones). (**C**) Measured transmission spectra of the two bands. (**D**) Optical picture of the color filter mosaic in transmission mode at the edge of the P-CFA (scale bar: 200μm). (**E**) Optical picture of the color filter mosaic in reflection mode near the edge of the P-CFA (scale bar: 200μm). (**F**) Normalized transmission spectra (after multiplication with sensor quantum efficiency) and cone spectral sensitivities, crosses showing the centroids of their corresponding spectra.

We used a set of Ishihara plates listed in Table 2 to validate the results. The plates were captured using the protanopic image sensor, and raw images were processed through the optimized neural network. Then we compared the output images against software-based protanopia simulations generated by Affinity Photo 2. The original Ishihara plates, reconstructed images from the neural network, and software-simulated images were all densely sampled and plotted onto the CIE Chromaticity Diagram (Table 3).

**Table 2.** Ishihara plates used for testing and their normal view & protanopic view ("Ishihara Charts," n.d.; Ishihara, 1972)

| Ishihara Plate No. | Normal View | Protanopia/protanomaly |
|---|---|---|
| 3 | 6 | 5 |
| 9 | 74 | 21 |
| 10 | 2 | most people don't see anything or see something incorrect |
| 21 | nothing | 73 |
| 23 | 42 | 2 |

**Table 3.** Original Ishihara plates, reconstructed images, simulated images, and their colors shown on CIE diagram

| | Original Image | Reconstructed Image by Neural Network | Simulated Protanopia by Affinity Photo 2 |
|---|---|---|---|
| Ishihara Plate No. 3 | 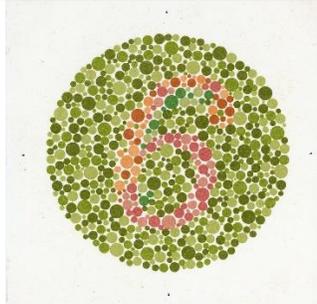 | 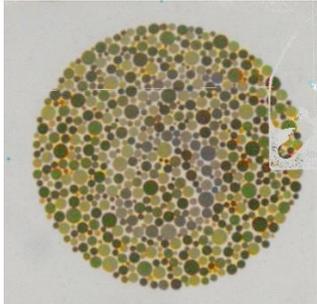 | 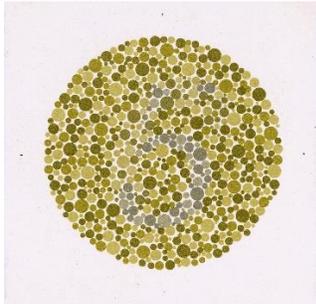 |
| CIE Diagrams | 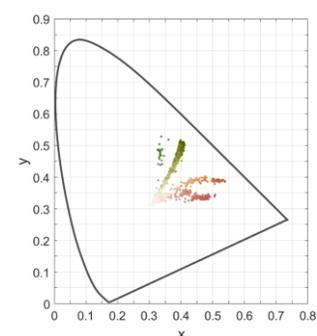 | 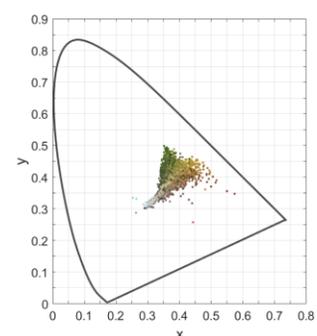 | 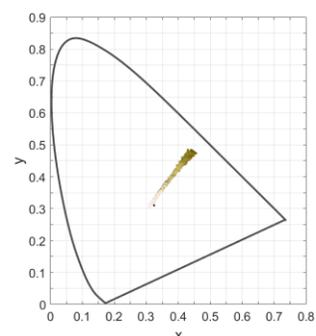 |
| Ishihara Plate No. 9 | 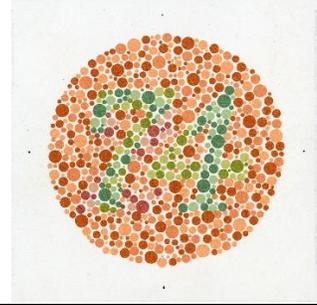 | 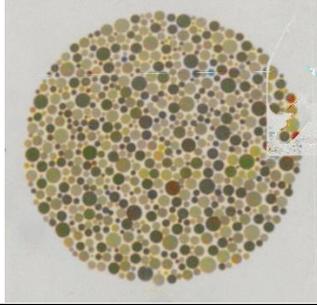 | 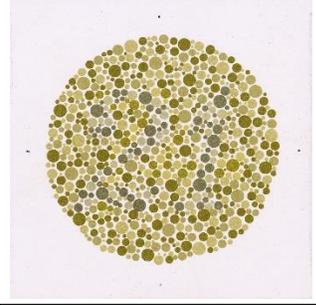 |
| CIE Diagrams | 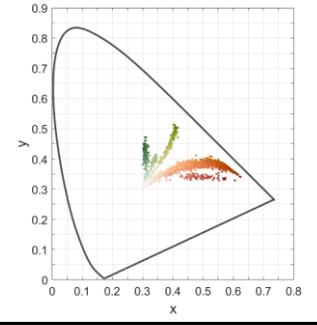 | 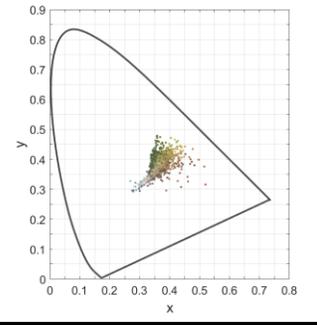 | 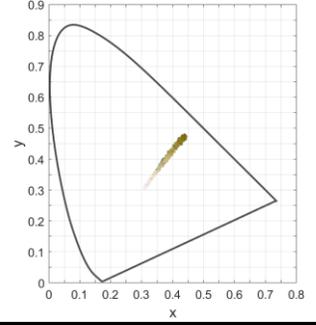 |
| Ishihara Plate No. 10 | 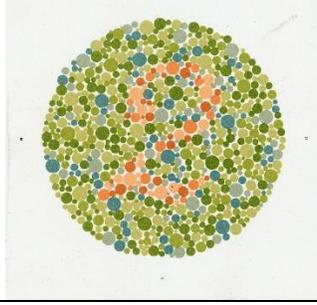 | 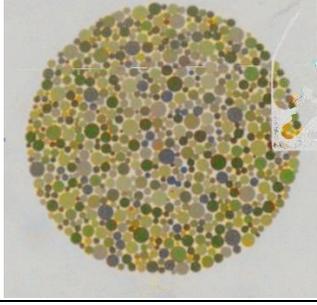 | 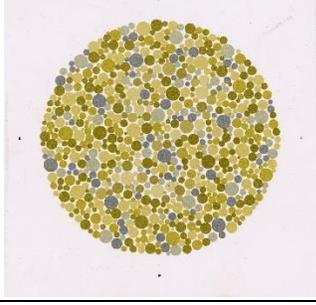 |
| CIE Diagrams | 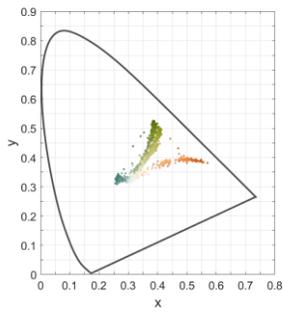 | 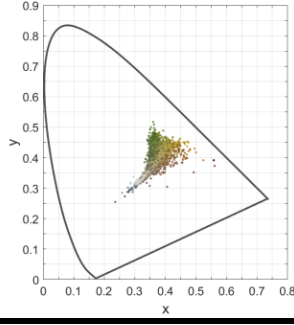 | 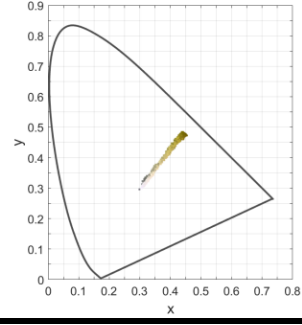 |
| Ishihara Plate No. 21 | 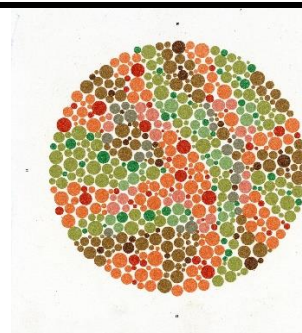 | 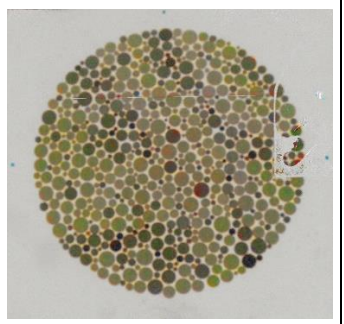 | 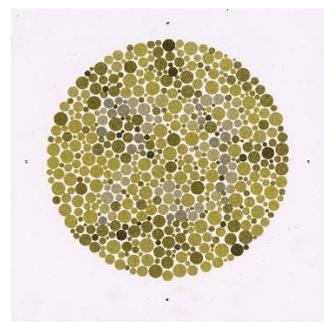 |
| CIE Diagrams | 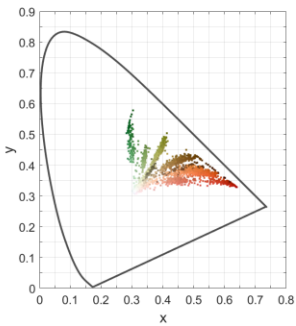 | 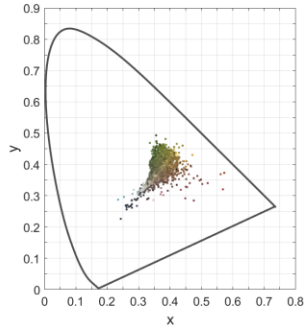 | 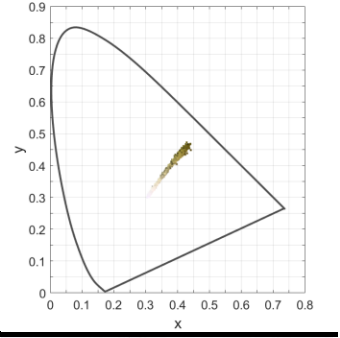 |
| Ishihara Plate No. 23 | 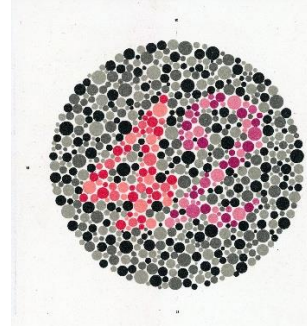 | 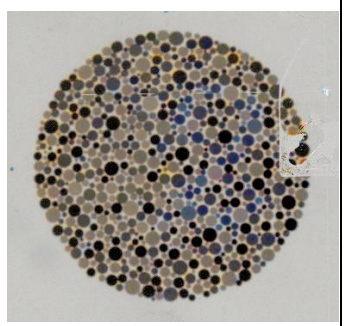 | 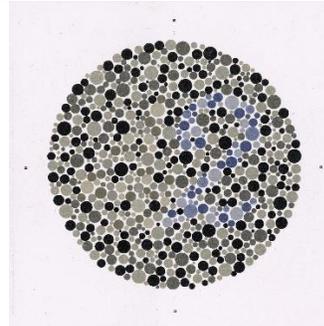 |
| CIE Diagrams | 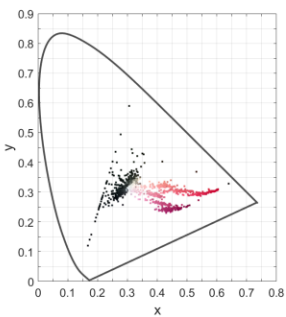 | 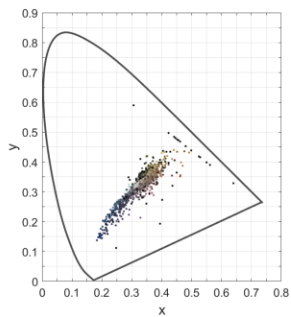 | 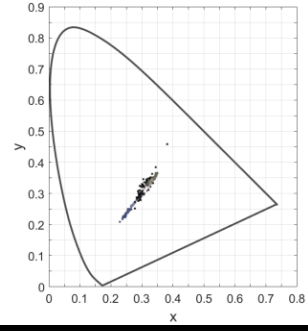 |

From the images in Table 3, it is evident that for Ishihara plate No. 3, 9 and 21, where individuals with protanopia or protanomaly should perceive specific numerals (Table 2), faint number 5, 21 and 73 appear in the reconstructed images. Likewise, for plate No. 10, the reconstructed image does not display any pattern as expected. For plate No. 23, according to the Ishihara book ("Ishihara Charts," n.d.; Ishihara, 1972), people with normal color vision read the plate as '42', however, those with protanopia or strong protanomaly distinguish '2' only, and those with mild protanomaly can see both numerals yet with '2' appearing clearer than '4'. The reconstructed image of plate No. 23 prominently displays a blueish number '2', and the reddish components have almost nearly vanished on the CIE diagram. The CIE diagrams also reveal that while pure protanopia narrows the color gamut to straight lines, our camera demonstrates semi-triangle shapes, which is consistent with the color range detectable by individuals with protanomaly (Fig. S8). This further confirms that our camera acts as a red-green color deficiency closer to protanopia and protanomaly. Therefore, this new color-blind camera effectively emulates protanopic-prone vision, as validated by established computational models and Ishihara books.

**Discussion**

In conclusion, this paper has presented an eye-inspired image sensor for actual implementation of personalized color vision deficiencies. A real foveal mosaic was adapted for the design of the Q-CFA, which resembles both the quasi-random distribution of L, M and S cone photoreceptors and their spectral sensitivities. The complete Q-CFA was integrated onto a 20.48Mpixel monochrome CMOS image sensor. In order to solve the random demosaicing problem, we employed a fully convolutional U-Net neural network that borrows ideas from human color vision by incorporating hierarchical convolutional layers to achieve pattern recognition through the receptive fields of ganglion cells. The training dataset was produced using a hybrid generation method, allowing the neural network to directly convert the raw RGBM captured mosaics into demosaiced sRGB color space images. It has been verified using the Macbeth ColorChecker® and Ishihara plates that our neural network is capable of both reconstructing full-color images as well as preserving the encoded hidden patterns in Ishihara charts. To prove the feasibility of our eye-inspired image sensor in studying color vision deficiencies, we developed a protanopic color blind image sensor by creating a new P-CFA that emulates the full human eye foveal mosaic as well as the spectral sensitivities of M and S cones. The neural network has been further optimized to produce images of Ishihara plates that exhibit desired colors and patterns in protanopic view as described in the official Ishihara books. By plotting colors of the reconstructed images onto the CIE chromaticity Diagram, we also validated that the camera has been successfully configured to mimic human eyes with a red-green deficiency closer to protanopia and protanomaly.

While largely successful, the eye-inspired image sensor based on the Q-CFA is still limited by its color perception function in a high but constant resolution, as the Q-CFA on the sensor is a repeated pattern of a small foveal mosaic instead of a full fovea patch. Although the P-CFA takes a further step by almost covering the entire central fovea region (7.4836×5 degrees), we have not fully exploited the vast potential of the distribution, which restricts the camera's ability to exhibit fully 'eye-like' behavior. Future studies should investigate the variation in cone size and density from the foveal center to the periphery, and develop compatible image processing algorithms to accommodate this variation, along with other effects such as eye movement (e.g., saccades). In addition, in this work the peak wavelengths of the three colored channels (Q-CFA) were shifted slightly from those of typical cones, especially in the case of the red filter with the purpose of reducing neural network complexity. While this approach helped to achieve the eye-inspired image sensor in our initial trial, it limits the ability of the mosaic to perfectly emulate the retina and will need to be addressed in future versions of the image processing algorithms. Moreover, as a result of limitations in the fabrication precision of the color filter arrays plus pixel alignment errors between the filter and sensor, neighboring pixels exhibit crosstalk between them, producing noise in the output images and reducing the color accuracy of some of the smaller circles in the Ishihara plates. To alleviate this noise issue, the current algorithm removes those pixels exhibiting severe crosstalk from the raw captured data and replaces them with values computed by the neural network. While this solution slightly compromises the image resolution, it proved to be more than adequate to support the development of our color-blind image sensor prototypes.

The ability to generate personalized mosaics from individual's full real retinal distributions and sensitivity functions is a promising future extension of this work and will enable further insights into color vision deficiencies by developing a digital twin of the human retina. This will open new ways to find accurate solutions for color blindness by capturing different images using the personalized color filter mosaic followed by developing suitable color filter solutions rather than using generalized color blindness solution, which is not developed based on individual's color deficiency mosaic. In addition, there is evidence suggesting that individuals with specific types of color vision deficiency may be more adept at recognizing certain patterns such as high frequency patterns, textures, or camouflage in nature. Our system can be employed to explore new sensing technologies aimed at these applications. Furthermore, our work will enable making new compact sensors combined with lidar to image high frequency patterns in 3D for applications in space and industry, including autonomous driving (Kamalakar et al., 2005; Liu et al., 2023).

**Materials and Methods**
    **Fabrication of the quasi-random color filter mosaic and protanopic color filter mosaic**

The fabrication of the Q-CFA is a three-stage fabrication process demonstrated in Fig. S2D. The color filter mosaic was built on a 4-inch circular glass wafer. The wafer was first spin-coated with AZ 2035 negative photoresist at 3000 rpm for 30 seconds, followed by soft-baking at 110°C for 2 min. Subsequently, the wafer was placed under UV exposure at 100 mJ/cm² using vacuum contact with the first photomask (blue channel), after which the wafer was post-baked at 110°C for 90 seconds, followed by development in AZ 726MIF developer for 45 seconds. Prior to the thin film depositions, the wafer underwent plasma pre-cleaning for 30 seconds. Following the layer sequence outlined in Table S1, $TiO_2$ and Ag/Au were deposited sequentially at constant deposition rates of 1Å/s and 0.3 Å/s, respectively. After the depositions, lift-off was performed by soaking the wafer in acetone for 1 min, followed by ultrasonic cleaning for 30 seconds. The wafer was then rinsed with acetone, isopropyl alcohol (IPA) and deionized (DI) water. The process involving photolithography, deposition, and lift-off steps was repeated for each mask step, with mask alignment being performed between each step. The fabrication of all three channels of the Q-CFA was completed after three iterations. The fabrication of both two channels of the P-CFA was completed after 2 iterations following the layer sequence outlined in Table S1.

**Spectrum measurement**

The optical spectra of the color filters were measured using a Cytoviva Hyperspectral Microscope. The wafer was first diced into 2748×1836 pixel-sized pieces, and a sample from the center area of the wafer was mounted on a glass slide under the microscope for observation at 50X magnification. For each of the three channels, 20 test points were randomly selected and approximately uniformly distributed across the sample. The measured spectrum for each channel corresponds to a normalized transmission result obtained by averaging the data from the 20 test points and dividing it by the spectrum of the glass slide alone.

**Hardware integration of the eye inspired image sensor**

The integration of the Q-CFA with the IMX183 CMOS monochrome image sensor involved careful pixel alignment using a FiconTech photonic assembly machine. During the alignment, the camera module (with cover glass removed from the IMX183 CMOS monochrome image sensor) was installed on the chuck at the bottom of the assembly machine while the filter piece was held by a side pick-up vacuum tool. The features on the filter and sensor were observed through the Topview vision system. Upon successful alignment between the filter pixels and sensor pixels, the filter piece was fixed onto the sensor with UV curable epoxy (further discussion on the alignment accuracy between the color filter array pixels and sensor pixels explained in supplementary information Section 8). The lens (V0826-MPZ, Computar), 1" type 20MP C-Mount with 8mm focal length, was attached to the camera. An additional optical bandpass filter (Hoya UV&IR Cut, Kenko Tokina Co., Ltd) was also added to remove UV below 390nm and IR above 700nm.

## Acknowledgments

Portions of this work were conducted in the Minnesota Nano Center, which is supported by the National Science Foundation through the National Nanotechnology Coordinated Infrastructure (NNCI) under Award Number ECCS-2025124.

This work was performed in part at the Melbourne Centre for Nanofabrication (MCN) in the Victorian Node of the Australian National Fabrication Facility (ANFF).

**Author contributions:**
- Conceptualization: YM, BW, PB, RU
- Fabrication and system integration: YM, DS, BW
- Experimental design and data analysis: YM, BW
- Supervision: RU, PB, PvW, ES, AN
- Writing—original draft: YM, BW, PB, RU
- Writing—review & editing: YM, BW, DS, PB, PvW, ES, AN, RU

**Competing interests:** Authors declare that they have no competing interests.

**Data and materials availability:** All data are available in the main text or the supplementary materials. For code, please contact the corresponding authors at r.ranjith@unimelb.edu.au, menym@student.unimelb.edu.au.